\documentclass{ws-rv961x669}
\usepackage{ws-rv-van}     
\usepackage{ws-rv-thm}     
\usepackage{subfigure}     
\usepackage{bm}
\makeindex

\begin{document}

\chapter[In the Pursuit of Majorana Modes in Iron-based High-$T_c$ Superconductors]{In the Pursuit of Majorana Modes in Iron-based High-$T_c$ Superconductors}\label{ra_ch1}

\author[Xianxin Wu, Rui-Xing Zhang, Gang Xu, Jiangping Hu and Chao-Xing Liu]{Xianxin Wu$^{1,2}$, Rui-Xing Zhang$^3$, Gang Xu$^{4,5}$, Jiangping Hu$^{2,6,7,\dag}$ and Chao-Xing Liu$^{1,*}$}  
\address{$^1$Department of Physics, the Pennsylvania State University, University Park, PA, 16802\\
$^2$Beijing National Laboratory for Condensed Matter Physics, and Institute of Physics, Chinese Academy of Sciences, Beijing 100190, China\\
 $^3$Condensed Matter Theory Center and Joint Quantum Institute, Department of Physics, University of Maryland, College Park, Maryland 20742-4111, USA\\
$^4$School of Physics, Huazhong University of Science and Technology, Wuhan, Hubei 430074, China\\
$^5$Wuhan National High Magnetic Field Center, Huazhong University of Science and Technology, Wuhan, Hubei 430074, China\\
$^6$Kavli Institute of Theoretical Sciences, University of Chinese Academy of Sciences, Beijing, 100049, China\\
$^7$Collaborative Innovation Center of Quantum Matter, Beijing 100049, China\\
jphu@iphy.ac.cn,cxl56@psu.edu} 

\begin{abstract}
Majorana zero mode is an exotic quasi-particle excitation with non-Abelian statistics in topological superconductor systems, and can serve as the cornerstone for topological quantum computation, a new type of fault-tolerant quantum computation architecture. This review paper highlights recent progress in realizing Majorana modes in iron-based high-temperature superconductors. We begin with the discussion on topological aspect of electronic band structures in iron-based superconductor compounds. Then we focus on several concrete proposals for Majorana modes, including the Majorana zero modes inside the vortex core on the surface of Fe(Te,Se), helical Majorana modes at the hinge of Fe(Te,Se), the Majorana zero modes at the corner of the Fe(Te,Se)/FeTe heterostructure or the monolayer Fe(Te,Se) under an in-plane magnetic field. We also review the current experimental stage and provide the perspective and outlook for this rapidly developing field.
\end{abstract}


\body


\section{Introduction}\label{sec:introduction}
In conventional solid materials, quasi-particle excitations normally obey either bosonic or fermionic statistics and their wavefunctions will acquire a phase factor of $e^{i\theta}$ ($\theta=0$ for bosons and $\theta=\pi$ for fermions) upon the exchange (braiding) of two quasi-particles. In two dimensions, collective excitations can in principle be anyons with more exotic statistical behaviors that are fundamentally different from bosonic or fermionic statistics \cite{stern2008anyons,wilczek1990fractional}. Exchanging two anyons can give rise to a fractional phase $\theta=\pi p/q$ with integers $p$ and $q$ for Abelian anyons or a unitary transformation among multiple degenerate many-body ground states for non-Abelian anyons\cite{wen1991non,Moore1991,read2000paired,ivanov2001non,kitaev2003fault,nayak2008non}. In particular, non-Abelian anyons can serve as topological quantum-bits (qubits) and the logic gate operations for these qubits can be implemented by simply braiding these anyons.\cite{kitaev2003fault,nayak2008non,aasen2016milestones,alicea2012new,beenakker2013search,sarma2015majorana}. The non-locality and topological nature of non-Abelian anyons can protect the stored quantum information against weak local perturbations from environments, thus providing a unique route to overcome the major challenge in quantum computation.

Majorana zero mode (MZM) is one example of such quasiparticle excitations, corresponding to a particular type of non-Abelian anyons called "Ising anyons". Here the zero mode refers to the zero-energy midgap excitation which appears inside the superconductor (SC) gap and normally locates at the boundary or inside the vortex core of low-dimensional topological SC (TSC) systems, including 5/2 fractional quantum Hall state (which can be viewed as a TSC of composite fermions) \cite{read2000paired}, p-wave Sr$_2$RuO$_4$ SCs\cite{rice1995sr2ruo4,sarma2006proposal} and the heterostructures consisting of SCs and spin-orbit coupled materials \cite{lutchyn2018majorana,oreg2010helical,Lutchyn2010,Sau2010,choy2011majorana,Nadj-Perge2014,Mourik2012,rokhinson2012fractional,Qi2010,fu2008superconducting,WangMX2012,ZhangH2018,alicea2010majorana,deng2016majorana,he2017chiral,kayyalha2019non,pientka2017topological,qi2011topological,hasan2010colloquium,bernevig2013topological}. Current major experimental efforts focus on the heterostructure approach in a variety of different systems, including semiconductor nanowires in proximity to SCs under magnetic fields
\cite{oreg2010helical,lutchyn2018majorana,Lutchyn2010,Sau2010,alicea2010majorana,deng2016majorana,Mourik2012,ZhangH2018}, magnetic ion chains on top of SC substrates \cite{Nadj-Perge2014,choy2011majorana}, the surface of topological insulator (TI) compound (Bi, Sb)$_2$Te$_3$ in proximity to SCs \cite{WangMX2012,fu2008superconducting,qi2011topological,hasan2010colloquium,bernevig2013topological} and the heterostructure with a quantum anomalous Hall insulator coupled to a SC \cite{Qi2010,he2017chiral,kayyalha2019non}. Despite of impressive experimental progress in the heterostructure approach, unambiguous detection and manipulation of MZMs in these heterostructures, however, heavily rely on the SC proximity effect that suffers from the complexity of the interface. Furthermore, the low operation temperature of conventional SC materials further complicates both the realization and manipulation of MZMs. On the other hand, high-temperature SCs, such as cuprate SCs\cite{bednorz1986possible} and iron-based SCs\cite{kamihara2008iron}, are not suitable for the heterostructure approach due to their short coherence length. It is thus desirable to find an intrinsic, robust and controllable Majorana platform that is compatible with existing fabrication and patterning technologies.

During the past few years, significant progress has been made in the search of intrinsic superconductor compounds that host non-trivial topological band structures, termed as ``{\it connate} TSCs"\cite{HaoNSR2019}. These compounds include Cu-, Nb- or Sr-doped Bi$_2$Se$_3$\cite{kriener2011bulk,asaba2017rotational,liu2015superconductivity}, p-doped TlBiTe$_2$\cite{yan2010theoretical,chen2010single,lin2010single}, p-doped Bi$_2$Te$_3$ under pressure\cite{zhang2011pressure}, half-Heusler SCs\cite{kim2018beyond,yan2014half,bay2012superconductivity,butch2011superconductivity}, and iron-based SCs\cite{HaoNSR2019}.

Among these SC compounds, iron-based SC family is of particular interest because of their high transition temperature, the abundance of the compounds in this family, and the intricate phase diagram with superconducting, nematic and magnetic orders. The earliest theoretical proposal in this direction was started by one of us for monolayer FeSe/STO\cite{Hao2014}.  Since then,  quite a few non-trivial band topology structures have been either theoretically proposed or experimentally observed in the normal states of CaFeAs$_2$ \cite{Wu2014PRB}, monolayer or bulk Fe(Te,Se) (FTS) \cite{Hao2014,Wu2016,Wang2015,Zhang2018}, Li(Fe,Co)As \cite{Zhang2019NP}, (Li$_{1-x}$Fe$_x$)OHFeSe \cite{PhysRevX.8.041056} and CaKFe$_4$As$_4$ \cite{2019arXiv190700904L}. These iron-based SC compounds provide ideal high-temperature SC platforms to explore Majorana modes and the relevant topological physics. In this article, we will review recent theoretical and experimental progress towards realizing Majorana physics in iron-based SCs. In Sec. \ref{sec:topological}, we will first discuss the underlying physical mechanism of topological electronic band structure in iron-based SCs. Sec. \ref{sec:Majorana} will review several theoretical proposals to achieve MZMs at different locations, including the vortex cores, the hinge and the corner, in iron-based SC systems, as well as recent experimental progress. Conclusion and outlook are given in Sec. \ref{sec:perspective}.

\section{Topological Band Structure in Iron-based superconductors}\label{sec:topological}
\begin{figure*}[tb]
\centerline{\includegraphics[width=\columnwidth]{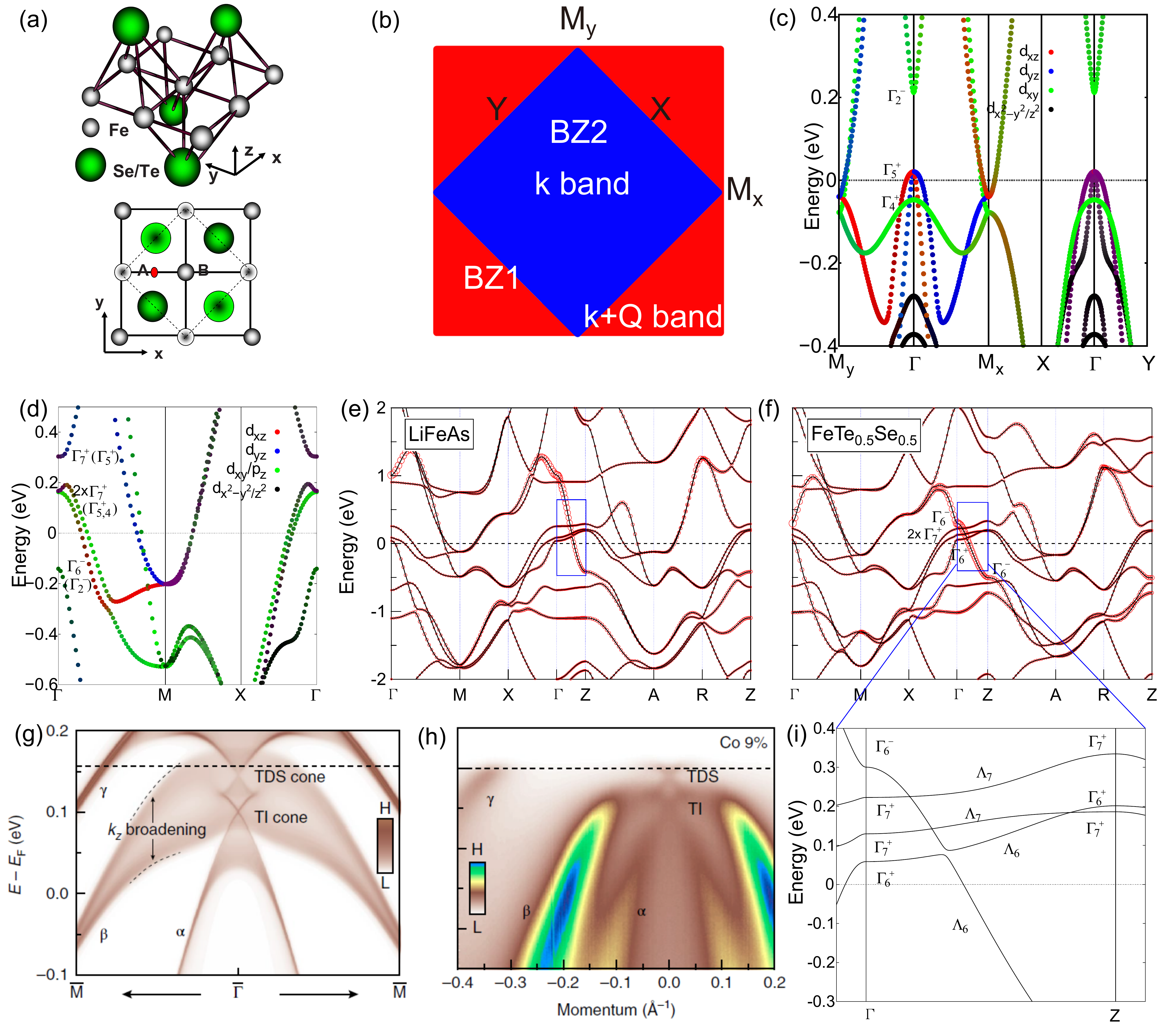}}
\caption{(color online) (a) Crystal structure for Fe(Te,Se) (Adapted from Ref. \cite{Hao2014}). (b) One-Fe BZ and two-Fe BZ. (c) The typical orbital-resolved band structure near the Fermi level for FeSe. (d) The inverted bands for monolayer Fe(Te,Se) in DFT calculations. Band structures for bulk LiFeAs (e) and FeTe$_{0.5}$Se$_{0.5}$ (f) and zoom-in band structure along $\Gamma$-Z (i) for FeTe$_{0.5}$Se$_{0.5}$ \cite{Zhang2019NP}. The area of red circles represent the $p_z$ orbital weight for As/Te,Se atoms. (001) surface spectrum of Li(Fe$_{1-x}$Co$_x$)As: (g) calculated surface spectrum; (h) ARPES intensity plot of Li(Fe$_{1-x}$Co$_x$)As (x = 9\%) \cite{Zhang2019NP}. The observed two Dirac cones originate from topological insulator phase and topological Dirac semimetal phase.  \label{TIFe} }
\end{figure*}

For most iron-based SCs, the electronic band structure is mainly determined by the Fe-X (X=As,P,Se,Te) tri-layer, as shown in Fig.\ref{TIFe}(a), in which the primitive unit-cell contains two iron and two pnictogen/chalcogen atoms. Due to the existence of the glide symmetry (combination of half translation and reflection in the xy plane), two iron and pnictogen/chalcogen sites in the primitive unit cell can be related to each other. Consequently, the unit-cell can also be redefined to contain only one Fe and one X atoms based on the glide symmetry \cite{Patrick2008,Tomic2014}. Therefore, in literature, the Brillouin zone (BZ) of iron-based SCs can be either defined for the primitive unit-cell (blue color region in Fig.\ref{TIFe}(b)) or for the reduced unit-cell (the whole regions with both red and blue colors in Fig.\ref{TIFe}(b)). In this paper, we will always discuss the electronic band structure in the BZ for two-Fe primitive unit-cell.

The energy bands near the Fermi energy are mainly contributed from the d-orbitals of two Fe atoms. Fig.\ref{TIFe}(c) show the electronic band structure and atomic orbital projections for monolayer FeSe film, from which one can see that the hole pocket around $\Gamma$ points are mainly attributed from $d_{xz}$ and $d_{yz}$ orbitals, while the electron pockets around M are dominated by $d_{xz}$, $d_{yz}$ and $d_{xy}$ orbitals (here $x,y$ is along Fe-Fe direction). Due to the multiple Fermi pockets, the extended s-wave SC pairing with the opposite sign of pairing gap at the electron and hole pockets becomes possible and this topic has been discussed and reviewed in literature\cite{Stewart2011,Hirschfeld2011}. Besides the superconductivity, an intriguing feature of the band structure is that the valence bands at $\Gamma$ with $d_{xz}$, $d_{yz}$ and $d_{xy}$ orbitals are of even parity under inversion (denoted as $\Gamma^+_{4,5}$ bands in Fig.\ref{TIFe}(c)) while the conduction band at $\Gamma$ is of $d_{xy}$ orbital character with odd parity (denoted as $\Gamma^-_2$ bands in Fig.\ref{TIFe}(c)). Therefore, according to the Fu-Kane criterion\cite{Fu2007PRB}, if the odd-parity $\Gamma^-_2$ bands drop below the even-parity $\Gamma^+_{4,5}$ bands, the system will undergo a topological phase transition and becomes topologically nontrivial. However, the conduction band bottom at $\Gamma$ has relatively high energy, around $0.2$ eV, above the valence band top. Furthermore, spin-orbit coupling (SOC) is negligible for the d orbitals of Fe atoms. Therefore, at the first sight, it seems hopeless to realize topological states in iron-based SCs.

A crucial theoretical insight comes from the understanding that the energy of the conduction band bottom strongly relies on the hybridization between the $d_{xy}$ orbital of Fe atoms and the $p_z$ orbital of anions (X atoms). Although the $p_z$-orbital band of anions is around $\sim 2$ eV below the Fermi energy, its parity (the bonding states of $p_z$ orbitals between two X-atom sites) is also odd and thus it can hybridize with the odd-parity $\Gamma^-_2$ bands, but not the even-parity $\Gamma^+_{4,5}$ bands at the $\Gamma$ point. Due to the level repulsion, this strong hybridization push the $\Gamma^-_2$ band above the $\Gamma^+_{4,5}$ bands in FeSe films. Therefore, by reducing the hybridization through enlarging the anion height with respect to Fe plane, the $\Gamma^-_2$ band bottom can sink in energy and eventually have lower energy than the $\Gamma^+_{4,5}$ bands, leading to a band inversion. It was theoretically proposed to enlarge the anion height by either replacing the Se atoms by Te atoms in FTS films or the strain effect from the substrate\cite{Wu2016}. Additional effect from the hybridization is to enhance SOC in the bands near the Fermi energy since the SOC strength of $p$-orbital of anions is normally significant. After including SOC, the $\Gamma_2^-$ ($\Gamma^+_4$) band should be changed to the $\Gamma_6^-$ ($\Gamma_7^+$) bands while the $\Gamma^+_5$ bands are split into the $\Gamma_6^+$ and $\Gamma_7^+$ bands. The new $\Gamma_6^-$, $\Gamma_6^+$ and $\Gamma_7^+$ bands are all spin degenerate due to the inversion and time reversal symmetry. The typical DFT band structure for FTS with inverted bands is shown in Fig.\ref{TIFe}(d), where the $\Gamma_6^-$ state sinks below the $\Gamma_{6,7}^+$ states. Since this band inversion mechanism only involves the energy bands around $\Gamma$, it can be captured by the well-established Bernevig-Hughes-Zhang (BHZ) model of the quantum spin Hall (QSH) effect\cite{Bernevig2006} on the basis of the odd-parity $\Gamma^-_6$ bands with the angular momentum $j_z=\pm\frac{1}{2}$ ($d_{xy}$ and $p_z$ orbitals) and the even-parity $\Gamma^+_7$ bands with the angular momentum $j_z=\pm\frac{3}{2}$ ($d_{xz}$ and $d_{yz}$ orbitals).
A more sophisticated study on this band inversion mechanism can also be carried out on a tight-binding model with all d-orbitals of Fe atoms and p-orbitals of X atoms, and justify this physical picture\cite{Wu2016}.

Similar physical mechanism also occurs in three dimensional (3d) bulk iron-based SCs. Different from two dimensional (2d) monolayer FTS, the $\Gamma_6^-$ energy bands are highly dispersive along the $k_z$ direction due to the $p_z$ orbital nature of X atoms. Fig.\ref{TIFe} (e) and (f) show the electronic band structure for FeTe$_{0.5}$Se$_{0.5}$ and LiFeAs, respectively. Fig.\ref{TIFe}(i) shows the zoom-in of the bands near the Fermi energy for FeTe$_{0.5}$Se$_{0.5}$ along the $\Gamma-Z$ line, from which one can see that the $\Gamma_6^-$ bands are above the $\Gamma_6^+$ and $\Gamma_7^+$ bands at $\Gamma$, but below these bands at $Z$ point in the 3d BZ. An anti-crossing gap is opened between the $\Gamma_6^-$ and $\Gamma_6^+$ bands along the $\Gamma-Z$ line, and leads to nontrivial topological state with surface states within this anti-crossing gap \cite{Wang2015}.
Fig.\ref{TIFe}(g) shows the direct theoretical calculations of topological surface states coexisting with bulk bands in Li(Fe$_{1-x}$Co$_x$)As, which shares a similar bulk band structure as FeTe$_{0.5}$Se$_{0.5}$ and LiFeAs. It should be mentioned that although the anti-crossing gap between the $\Gamma_6^-$ and $\Gamma_6^+$ bands is $\sim 0.1$eV above the Fermi energy in the DFT calculation, the ARPES measurements show that the Fermi energy exactly lies within this gap for FeTe$_{0.5}$Se$_{0.5}$ and LiFeAs, thus allowing for the direct observation of the topological surface states \cite{Zhang2018,Zhang2019NP}. This discrepancy presumably comes from the inadequate treatment of the strong correlation effect of iron-based SCs in the DFT calculations. In addition to the anti-crossing between the $\Gamma_6^-$ and $\Gamma_6^+$ bands, the band sequence between $\Gamma^-_6$ and $\Gamma^+_7$ bands is also inverted, but these two bands can cross with each other along the $\Gamma$-$Z$ path and form a 3d Dirac cone protected by the $C_4$ rotational symmetry due to the different eigen-values of $C_4$ rotation for these two bands ($e^{\pm \frac{\pi}{4}i}$ for $\Gamma^-_6$ bands and $e^{\pm\frac{3\pi}{4}i}$ for $\Gamma^+_7$ bands). A complete eight-band tight-binding model including $\Gamma_6^-$, $\Gamma^+_6$ and two $\Gamma^+_7$ bands has been developed by Xu {\it et al} \cite{XuPRL2016} and all the relevant topological physics of this system can be studied within this model. Further simplified models also exist. The non-trivial anti-crossing between the $\Gamma_6^-$ and $\Gamma_6^+$ bands and the topological surface states can be studied within a four-band model that was first developed for the prototype TI Bi$_2$(Se,Te)$_3$ family of materials \cite{zhang2009topological,liu2010model}. On the other hand, the crossing between $\Gamma^-_6$ and $\Gamma^+_7$ bands can be well described by the 3d Dirac Hamiltonian which was first used for 3d Dirac semimetals, such as Na$_3$Bi \cite{WangZJPRB2012} and Cd$_2$As$_3$ \cite{WangZJPRB2013}.

It should be emphasized that the above physical scenario is not limited for one or two specified compounds, but generally exists in most families of iron-based SCs, including 122, 111 and 11 families\cite{Zhang2019NP}, thus making iron-based SCs really a fertile platform to explore TSC physics. The physics can be slightly different for different compounds. For example, in most iron-based SC compounds, both topological properties and high T$_c$ superconductivity originate from Fe-X (X=As,Se,Te) tri-layers, but CaFeAs$_2$ is an exception. In CaFeAs$_2$, the superconductivity comes from FeAs layer while the topologically non-trivial bands occur in the CaAs layer. Thus, CaFeAs$_2$ can be regarded as an intrinsic TI-SC hetero-structure\cite{Wu2014PRB}. Furthermore, the band inversion may also occur around M point, in additional to the $\Gamma$ point, as demonstrated in FeSe with the strain effect from the substrate.\cite{Hao2014}

Exciting experimental progress has been achieved for probing topological electronic bands of iron-based SCs during the past years. The angular resolved photoemission spectroscopy (ARPES) measurements with high energy and momentum resolution were performed for the (001) surface of FTS, which clearly resolves the surface Dirac cone \cite{Zhang2018}. Moreover, in the same experiment, spin-resolved ARPES measurements directly verify the helical nature of the surface states, confirming their topological origin. Additionally, in the iron pnictide Li(Fe$_{1-x}$Co$_x$)As \cite{Zhang2019NP}, the chemical potential can be controlled by tuning the Co doping ratio. Therefore, both the topological surface states within the anti-crossing gap between the $\Gamma_6^-$ and $\Gamma_6^+$ bands and the 3d bulk Dirac cone due to the crossing between the $\Gamma_6^-$ and $\Gamma_7^+$ bands have been simultaneously observed, as shown in Fig.\ref{TIFe}(h), in good agreement with the theoretical calculations in Fig.\ref{TIFe}(g).

For 2d FTS thin films, strong evidence for band inversion has been found in ARPES measurements when tuning the Se concentrations of FTS\cite{Shi2017}. Very recently, ARPES measurements reveal the topological phase transition process with the variation of Se concentration in monolayer FTS/STO, while in the samples with inverted bands, the scanning tunneling microscopy (STM) measurements provide strong evidence for the existence of nontrivial edge states.\cite{Peng2019}

\section{Majorana Modes in Iron-based superconductors}\label{sec:Majorana}
The generic coexistence of superconductivity and multiple topological states makes a broad class of iron-based SCs a promising high $T_c$ SC platform for exploring topological SCs and Majorana modes. The topics of Majorana physics in topological SCs has been reviewed in a number of papers \cite{alicea2012new,beenakker2013search,sarma2015majorana,lutchyn2018majorana}, to which the readers can refer for more details. Here we focus on how to achieve Majorana physics in different sample configurations based on iron-based SCs. In particular, it has been theoretically proposed that MZMs can exist inside the vortex cores or at the corners while the helical Majorana modes can appear at the hinge of iron-based SCs.\cite{XuPRL2016,zhang2019helical,zhang2019higher,wu2019high} The corner and hinge Majorana modes are also relevant to another emergent subfield, called higher-order topological state.

\subsection{Vortex line transition and Majorana zero modes in the vortex core}\label{sec:Majorana_vortex}
The early studies on MZMs focus on the intrinsic $p+ip$ superconductor, in which the MZMs can be localized in the $h/2e$ vortex core \cite{read2000paired} for the 2d case or at the end for the one dimensional (1d) case\cite{kitaev2003fault}. However, the intrinsic $p+ip$ superconductor is rare in nature and the experimental evidence of MZMs in p-wave SC, such as Sr$_2$RuO$_4$\cite{rice1995sr2ruo4,sarma2006proposal}, is still lacking. In a seminal work \cite{fu2008superconducting}, Fu and Kane notice that the effective Hamiltonian for the surface states of a strong TI in proximity to a s-wave SC is equivalent to that of a spinless $p+ip$ SC up to a unitary transformation. Consequently, the MZM is also expected to exist at the $h/2e$ vortex core at the surface of a TI with a conventional SC deposited on top. Although the original idea is for a TI-SC heterostructure, the iron-based SCs provide an ideal intrinsic platform to realize this theoretical proposal. As discussed above, the topological surface states have been experimentally observed in several iron-based SC compounds, with the example of FTS\cite{Zhang2018}. Besides the surface states, bulk electron and hole bands also exist at the Fermi energy and are responsible for the occurrence of superconductivity in these compounds. The intrinsic or self-proximity effect for the superconductivity from the bulk bands to the topological surface bands has also been demonstrated experimentally. For example, a superconducting gap around 1.8~meV has been observed on the topological surface state of FeTe$_{0.55}$Se$_{0.45}$ in the ARPES measurements~\cite{Zhang2018}.
However, there is still one subtle issue. Fu and Kane's original proposal relies on the Fermi energy only crossing the topological surface bands, while in iron-based SCs, the surface bands are normally buried in the bulk bands. Thus, it is natural to ask if the MZMs inside the vortex core at the surface can still survive when both bulk and surface bands appear at the Fermi energy and are strongly hybridized.

\begin{figure}
	\centering
\includegraphics[width=1.0\textwidth]{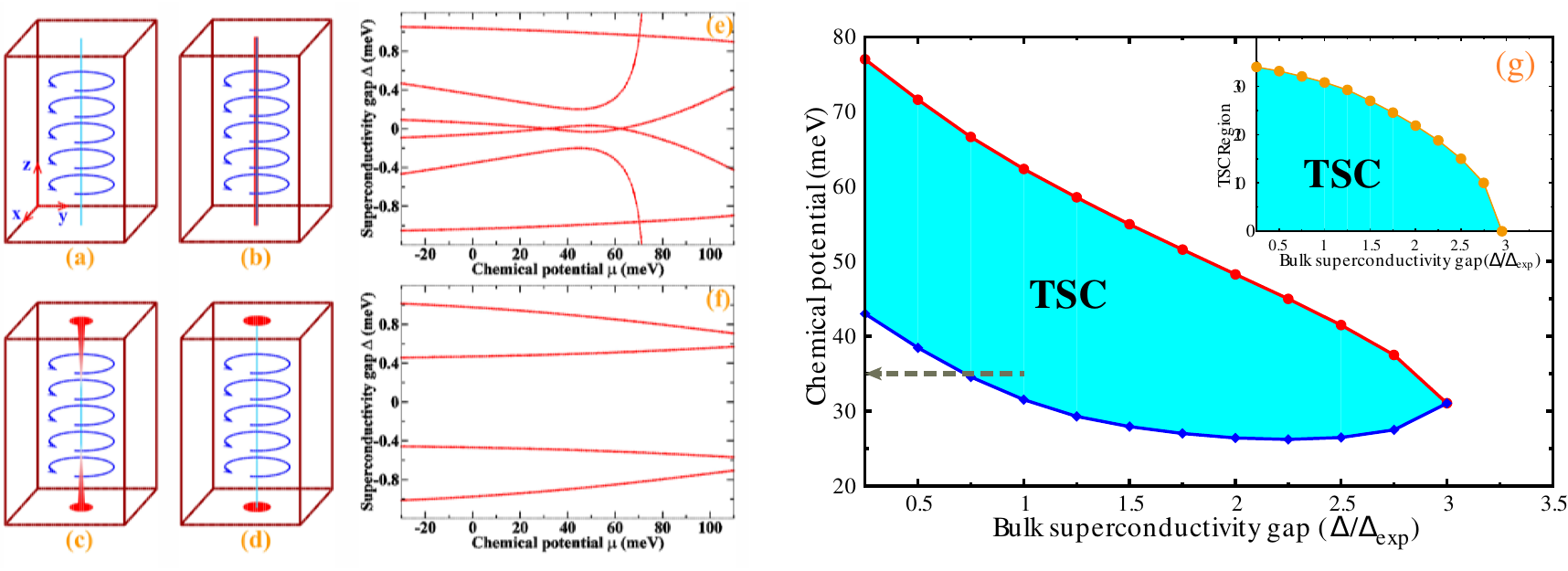} 
\caption{
(a) - (d) The schematic evolution of the surface MZMs in
a vortex line. As the chemical potential is tuned from the trivial
regime (a) towards the surface TSC regime, two Majorana zero
modes arise [(b)] and then become more and more localized at the
ends of the vortex line [(c) and (d)]. (e) The energy spectrum at
the $\Gamma$ point of a vortex line along the z direction as a function of
the chemical potential $\mu$. The energy gap closes at $\mu_1 = 31$~meV and
$\mu_2 = 62$~meV, respectively, showing the $(001)$ surface is a TSC in
the range $\mu \in (\mu_1, \mu_2)$. (f) The energy spectrum at the $Z$ point of
the vortex line as a function of the chemical potential $\mu$. There is
no gap closing at the $Z$ point, so the phase transitions are solely
determined by the gap closing at the $\Gamma$ point. In (a)-(f),
chemical potential $\mu = 0$ corresponds to the Fermi level of the
stoichiometric FST.
(g) TSC phase space vs bulk superconducting gap. The red
and blue lines are the upper and lower phase boundaries of the
TSC phase, respectively. The gray dash at $\mu = 35$~meV indicates the phase transition from
a TSC at 0 K ($\Delta = \Delta_{\text{exp}}$) to NSC at $T = T_c$ ($\Delta \sim 0$) with increasing temperature. The inset shows an evolution of the TSC region (red line minus blue line) with respect to the bulk
pairing gap. From Ref.~\cite{XuPRL2016}.
 }
\label{fig:vortexline}
\end{figure}

This question was first addressed by Hosur~\emph{et al} in a model for doped TIs, which can be applied to several SC compounds based on doped Bi$_2$(Se,Te)$_3$ family of materials \cite{Hosur2011}. It was found that there is a critical doping, at which a one dimensional (1d) topological phase transition can occur along the vortex line. This vortex line phase transition separates a trivial phase with a fully gapped vortex line from a non-trivial phase, in which MZMs are trapped inside the vortex core at the surface of doped TIs (or equivalently at the end of the vortex line), as shown in Fig. \ref{fig:vortexline}(a)-(d). Compared to doped TI systems, iron-based SCs possess a more complex electronic band structure near the chemical potential. Specifically, the $\Gamma_6^-$ bands are inverted with two $\Gamma_7^+$ and one $\Gamma_6^+$ bands along the $\Gamma-Z$ line, and all these bands may affect the existence of vortex core MZMs at the surface. Furthermore, it is well accepted that iron-based SCs have an extended s-wave pairing, rather than a simple s-wave pairing, and its influence of the surface MZMs is unclear.

To test this idea in bulk Fe$_{1+y}$Se$_{0.5}$Te$_{0.5}$, Xu~\emph{et al.}\cite{XuPRL2016} built up an eight-band tight-binding model including all the bands that are relevant for the band inversion ($\Gamma_6^-, \Gamma_6^+$ and two $\Gamma_7^+$ bands with spin degeneracy). Based on this model, the bulk energy dispersion can fit well with that from the first principles calculations and the topological surface states can be found on the (001) surface, thus confirming the non-trivial topology of this material. By further including the extended s-wave pairing into this model, Xu~\emph{et al.} numerically studied the energy levels of a magnetic vortex line in a cylinder geometry for Fe$_{1+y}$Se$_{0.5}$Te$_{0.5}$ with the momentum $k_z$ still being a good quantum number along the vortex line direction. Their numerical results show that the SC gap at $\Gamma$ point along the magnetic vortex line closes at two different chemical potentials $\mu_1$ and $\mu_2$ shown in Fig. \ref{fig:vortexline}(e), while it remains open at $Z$ point for all the $\mu$ values in Fig. \ref{fig:vortexline}(f). This suggests the vortex line transition occurring at two chemical potentials $\mu_1$ and $\mu_2$, and the system is in the TSC phase with the MZMs localized in the vortex core at the (001) surface when $\mu_1<\mu<\mu_2$. The phase diagram on the parameter space spanned by the pairing strength $\Delta$ and chemical potential $\mu$ is also obtained and depicted in Fig. \ref{fig:vortexline}(g), from which a finite regime of TSC phase is identified. From this phase diagram, a vortex line transition can be induced by varying temperature for a fixed $\mu$. For instance, at $\mu = 35$ meV, the system is in the TSC phase at zero temperature ($\Delta \sim \Delta_{\text{exp}}$), but becomes a trivial SC phase when temperature is close to $T_c$ ($\Delta \rightarrow 0$), as indicated by the gray dashed arrow in Fig. \ref{fig:vortexline}(g).

Recently, several experimental groups have reported the observation of zero bias peak at the vortex core at the surface of FeSe$_{0.45}$Te$_{0.55}$~\cite{Wang333, Machida2019, 2019arXiv190102293K},
and (Li$_{0.84}$Fe$_{0.16}$)OHFeSe~\cite{PhysRevX.8.041056} through STM measurements. The spatial profile and the magnetic field dependence of zero bias peak are consistent with the physical picture of MZM, thus providing strong evidence of its topological origin, while alternative explanation based on the conventional Caroli-de Gennes-Matricon states was still not excluded~\cite{ChenMingyang2018}.

Besides MZMs at the end of the vortex line, recent theoretical studies demonstrate that a quasi-1d helical Majorana state protected by $C_{4z}$ rotation symmetry can exist along the vortex line when the chemical potential is tuned close to the 3d Dirac cone formed by the $\Gamma_6^-$ and $\Gamma_7^+$ bands~\cite{KoenigPRL2019,PhysRevLett.123.027003,Kawakami2019}. When the bulk state is in the 3d weak TI phase, such as in (Li$_{1-x}$Fe$_x$)OHFeSe, Qin~\emph{et al.} showed that the vortex phase can also be nodal SC with pairs of helical Majorana modes along the vortex line~\cite{QIN20191207}.

\subsection{Higher-Order Topology in Iron-based Superconductors }\label{sec:Majorana_high_order}

The past few years have witnessed the rapid developments of ``higher-order topology" as a direct generalization of TIs \cite{benalcazar2017quantized,schindler2018higher}. By definition, a $D$-dimensional topological state has $n$th-order topology if the system has non-trivial gapless state on some of its $(D-n)$-dimensional boundary manifolds. For example, a 2nd-order TI in 3d is usually featured by gapped 3d bulk and 2d surface states, just like a topologically trivial atomic insulator. However, there exist 1d gapless modes that live on the ``hinges" between adjacent surfaces. Experimentally, if we simply probe a 2nd-order TI with ARPES measurement, we will most certainly conclude that it is a trivial band insulator given the gapped surface spectrum. The higher-order topology is thus only revealed when the sample hinges are probed by STM, transport studies and other hinge-sensitive measurements. In experiment, phenomena of higher-order topological insulators have been observed in photonic systems \cite{Peterson2018quantized}, phononic systems \cite{Serra-Garcia2018observation}, acoustic systems \cite{Xue2019acoustic} and bismuth \cite{Schindler2018bismuch}.

Compared with the higher-order TIs, their superconducting counterparts, the higher-order TSC, appear to be more exotic for hosting corner or hinge Majorana modes \cite{YanZB2018,WangQY2018,WangYX2018,zhu2018tunable,pan2018lattice}, which provide a new platform for topological quantum computation. One promising route to achieve higher-order TSC is to start with a normal state TI and further introduce bulk unconventional superconductivity. As an example, let us consider a 3d superconducting TI whose 2d Dirac surface states become gapped due to the pairing effects. When the bulk pairing is {\it not} isotropic s-wave, it is possible that the projected surface pairings will acquire a $\pi$-phase difference for two neighboring surfaces. In this case, the hinge between these two surfaces is equivalent to a superconducting domain wall of the surface Dirac fermions, which necessarily binds 1d Majorana modes. As a result, such superconducting TI system realizes 1d hinge Majorana modes along with gapped bulk and surfaces, which manifests itself as a second-order TSC in three dimensions. In principle, we can consider similar physics in a 2d superconducting TI that realizes Majorana zero modes localized on the sample corners. A prerequisite for this simple mechanism is the coexistence of nontrivial band topology and unconventional superconductivity, which has been experimentally demonstrated in a number of iron-based SC compounds. To demonstrate this idea of higher-order TSC phase in iron-based SCs, we will first describe how 1d dispersing helical Majorana modes naturally emerge on the hinges of a bulk FTS following Ref. \cite{zhang2019helical}. While a FTS monolayer is unlikely higher-order topological by itself, we will show that localized Majorana zero modes will appear on the sample corners, when the monolayer is covered by an additional FeTe monolayer \cite{zhang2019higher} or placed under in-plane magnetic field \cite{wu2019high}.

\subsubsection{1d Helical Majorana modes on the hinges}\label{sec:Majorana_hinge}
\begin{figure}[t]
	\centering
	\includegraphics[width=1.0\textwidth]{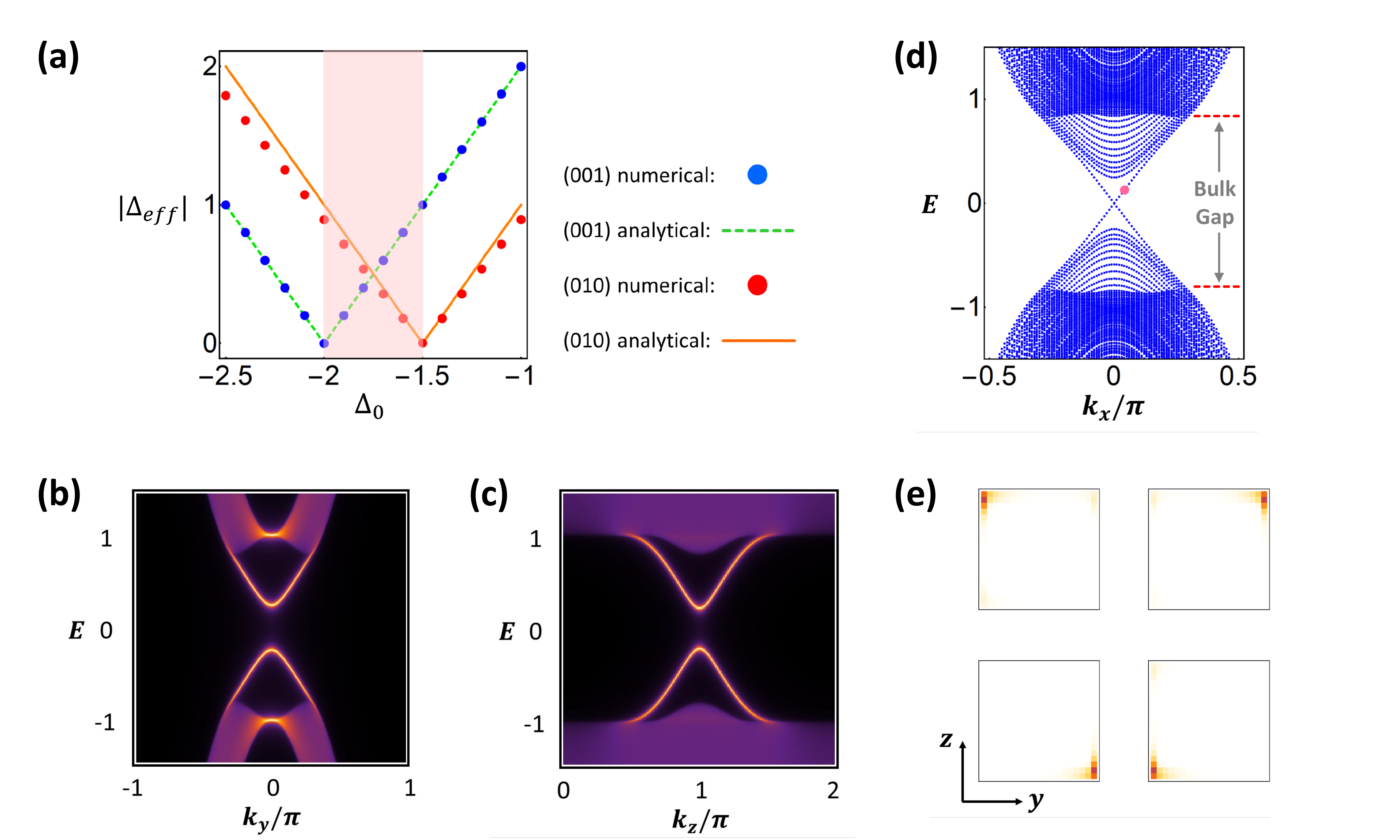}
	\caption{(a) Evolution of the (001) and (010) surface gaps. The green dashed and red solid lines use Eq. \ref{Eq: FTS surface pairing}. Results for both surfaces agree well with numerical calculations on the full lattice model (dots). The red region denotes the phase with helical hinge Majorana modes. In (b) and (c), we plot the surface spectrum of both (001) and (010) surfaces using the iterative Green function method. (d) Energy spectrum in a wire geometry along $x$ with open boundary conditions on both $y$ and $z$ with 20 lattice sites along each direction. The linear modes inside the surface gap are the helical hinge Majorana modes. (e) Spatial profile of the eigenstates at the red dot in (d). From Ref. \cite{zhang2019helical}.}
	\label{Fig: hinge}
\end{figure}

The idea that an iron-based superconductor can be {\it intrinsically} a higher-order TSC was first proposed and demonstrated by Zhang {\it et al.} in Ref. \cite{zhang2019helical}. This proposal is inspired by the ARPES observation of spin-momentum locked topological Dirac surface state in the normal state of bulk FeTe$_x$Se$_{1-x}$, which is also believed to develop the extended s-wave pairing (also known as $s_{\pm}$ pairing) below the superconductor transition temperature $T_c$. Being an unconventionally superconducting TI, the central question for FTS is whether hinge Majorana modes could show up as a result of the in-plane $s_{\pm}$ pairing.

To model the topological physics in FTS, Zhang {\it et al.} construct a minimal lattice model that captures the following important properties of this material: (i) a topological band inversion at $Z$ in the Brillouin zone; (ii) bulk $s_{\pm}$ pairing with $\Delta({\bf k}) = \Delta_0 + \Delta_1 (\cos k_x + \cos k_y)$. Here, $\Delta_0$ and $\Delta_1$ denote on-site and nearest-neighbor s-wave pairing, respectively.

To understand the origin of higher-order topology in FTS, Zhang {\it et al.} derived an effective surface theory $H_{\Sigma}$ in the continuum limit for an arbitrary surface $\Sigma(\phi,\theta)$. Here the surface $\Sigma(\phi,\theta)$ is characterized by a unit vector ${\bf n}=(\sin\theta \cos\phi, \sin\theta \sin \phi, \cos \theta)^T$ in the spherical coordinate, and can thus be understood as the tangent surface of a unit sphere defined by ${\bf n}$. By solving for the surface theory of $\Sigma(\phi,\theta)$ analytically and project the bulk $s_{\pm}$ pairing onto the surface state basis, Zhang {\it et al.} obtain the effective surface theory for $\Sigma(\phi,\theta)$ as
\begin{equation}
	H_{\Sigma} (k_1,k_2) = \begin{pmatrix}
	k_1 \varsigma_2 + k_2 \varsigma_1 & -i\varsigma_y\Delta_\text{eff}(\theta) \\
	i\varsigma_y\Delta_\text{eff}(\theta) & -k_1 \varsigma_2 + k_2 \varsigma_1 \\
	\end{pmatrix}.
\label{Eq: FTS surface theory}
\end{equation}
Here, the in-plane crystal momenta $(k_1,k_2)$ on the surface $\Sigma(\phi,\theta)$ is related to $(k_x,k_y)$ by two successive Euler rotations $R_Z(-\phi)$ and $R_Y(-\theta)$ around $z$ and $y$ axes, respectively. $\varsigma_i$ are defined as the Pauli matrices in the TI surface state bases. Importantly, the effective pairing gap on the surface has a strong $\theta$ dependence, which is given by
\begin{equation}
	\Delta_\text{eff} (\theta) = \Delta_0 + 2 \Delta_1 - \Delta_1 \frac{m_0-2m_1+m_2}{m_2\cos^2\theta-m_1\sin^2\theta}\sin^2\theta.
	\label{Eq: FTS surface pairing}
\end{equation}
where $m_{0,1,2}$ are model parameters. Note that $\Delta_\text{eff}(\theta)$ is isotropic in $\phi$, which simply because the continuum limit of $H({\bf k})$ has full rotational symmetry around $z$ axis. With Eq. \ref{Eq: FTS surface pairing}, Zhang {\it et al.} arrive at a simple criterion for higher-order TSC in FTS: {\it Given two surfaces with distinct values of $\theta$ (e.g. $\theta_1$ and $\theta_2$), the hinge connecting them will necessarily host a pair of helical Majorana modes if $\Delta(\theta_1)\Delta(\theta_2)<0$.} In particular, for the band parameters used in their work, the hinge Majorana physics emerges when $-2\Delta_1<\Delta_0<-1.5\Delta_1$. Physically, this topological criterion is the condition to form a superconducting mass domain wall betweem the two surfaces. Thus, the helical hinge Majorana modes can be interpreted as the domain wall modes that arise due to the $\pi$-phase difference of the effective surface pairing functions.

Zhang {\it et al.} further provide numerical results to demonstrate the higher-order topological physics in this model. As shown in Fig. \ref{Fig: hinge} (a), they first calculate the surface gap evolution as a function of $\Delta_0$ for both (001) and (010) surfaces. By comparing with the analytical results predicted by Eq. \ref{Eq: FTS surface pairing}, they found a quantitative agreement between these results. At $\Delta_0=-1.75\Delta_1$, they observed gapped energy spectra for both $(001)$ and $(010)$ surfaces in the semi-infinite geometry, as shown in Fig. \ref{Fig: hinge} (b) and (c). To reveal the hinge Majorana modes, they further performed a calculation for the energy spectrum in a infinitely long wire geometry along $x$ direction, with open boundary conditions imposed for both $y$ and $z$ directions. As shown in Fig. \ref{Fig: hinge} (d), while both bulk and surface gaps are clearly illustrated, there exist four pairs of 1d helical Majorana modes penetrating the surface gap. By plotting the spatial profile of these Majorana modes, Fig. \ref{Fig: hinge} (e) demonstrates the localization of these modes around four corners of the $y$-$z$ cross section. Consequently, for a cubic sample with open boundary conditions in all three spatial directions, the helical Majorana modes will propagate along the hinges between top/bottom and side surfaces, which thus establishes their nature as {\it helical hinge Majorana modes}.

Zhang {\it et al.} also studied the robustness of the higher-order topology in their model with respect to a finite chemical potential and weak potential disorder effects. As the first material proposal of 3d higher-order TSC, Ref. \cite{zhang2019helical} also inspires experimental efforts to probe the hinge Majorana physics in the iron-based materials. For example, Ref. \cite{gray2019evidence} performed a systematic point contact study on FTS samples and observed interesting zero-bais peaks in the tunneling spectroscopy that suggest possible Majorana-related physics on the sample hinges.

\subsubsection{Majorana zero modes at the corners}\label{sec:Majorana_corner}
Recently, monolayer FTS, a known 2d high-T$_c$ superconductor, is also found to host non-trivial band topology in its normal state.Specifically, topological band gap closing process at $\Gamma$ point has been observed through ARPES and STM techniques\cite{Peng2019}. Thus, the normal state physics resembles that of a 2d TI. Unlike its 3d counterpart, however, the pairing mechanism of monolayer FTS remains mysterious and under debate, which makes it difficult to concretely predict promising topological phenomenon based on this 2d platform. Despite the normal-state band topology, it is nonetheless unlikely for monolayer FTS to be topologically superconducting in an intrinsic way. Therefore, a natural question is that whether and how Majorana physics could emerge in monolayer FTS.

This challenge is resolved simultaneously by Zhang {\it et al.} in Ref. \cite{zhang2019higher} and Wu {\it et al.} in Ref. \cite{wu2019high}, where 2d higher-order TSC phases with corner Majorana modes are independently proposed by breaking the time-reversal symmetry externally. In particular, Zhang {\it et al.} proposed to reveal the higher-order topology with a hetero-structure combining monolayer FTS and FeTe with anti-ferromagnetic ordering, while Wu {\it et al.} claimed that applying an in-plane magnetic field to FTS could do the same job. In the following, we will review these two complimentary proposals for realizing Majorana corner modes in monolayer FTS.

\begin{figure}[t]
	\centering
	\includegraphics[width=1.0\textwidth]{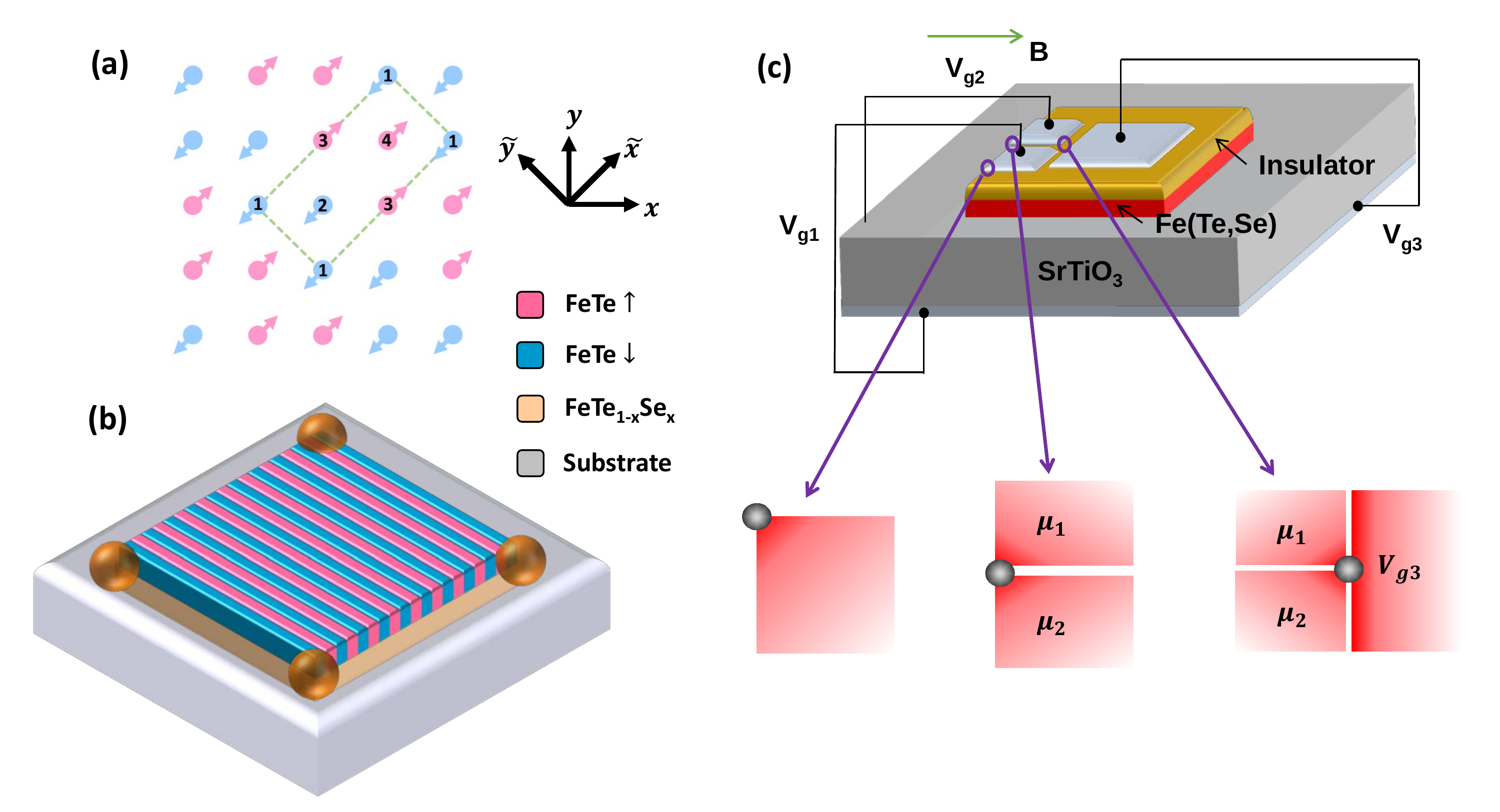}
	\caption{(a) Schematic plot of bicollinear antiferromagnetic order in FeTe. The circle and its arrow represent the Fe atom and its magnetic moment. (b) Schematic plot of the FTS/FeTe heterostructure with corner-localized Majorana modes. (c) Schematics for the Majorana platform based on monolayer FTS under the in-plane magnetic field. MZMs can be found at three different locations: the corner between two perpendicular edges, the chemical potential domain wall along the 1d edge and the
tri-junction in the 2d bulk. From Ref. \cite{zhang2019higher,wu2019high}.}
	\label{Fig: Corner modes schematic}
\end{figure}

The proposal from Zhang {\it et al.} takes advantageous of the abundant magnetic and superconducting physics in the iron chalcogenides.	In particular, the transition between an antiferromagnet (AFM) and a SC is simply tuned by the ratio between Se and Te in FTS. For example, FeSe and FeTe represent examples for a high-Tc superconductor and a bi-collinear AFM [see Fig. \ref{Fig: Corner modes schematic} (a) for the magnetic configuration], respectively. Therefore, by growing an extra FeTe layer on top on the existing FTS monolayer, the AFM structure of FeTe will compete with the superconductivity in FTS in an anisotropic way. When projected on the edges, this anisotropy turns out to be the key to drive the system into a higher-order TSC.

To demonstrate this idea, Zhang {\it et al.} considered the bilayer heterostructure that consists of a FTS monolayer and a FeTe monolayer, as shown in Fig. \ref{Fig: Corner modes schematic} (b). They wrote down a Bernevig-Hughes-Zhang (BHZ) model to capture the TI nature of monolayer FTS around $\Gamma$ point of the Brillouin zone, following Ref.\cite{Wu2016}. Introducing bicolinear AFM order $M$ and singlet s-wave pairing $\Delta$ enlarges the unit cell to include four inequivalent atom sites, which is shown by the green dashed line in Fig. \ref{Fig: Corner modes schematic} (a). The lattice vectors for the enlarged unit cell $\tilde{\bf a}_{x,y}$ are related to the original lattice vectors ${\bf a}_{x.y}$ as $\tilde{\bf a}_x = 2({\bf a}_x + {\bf a}_y)$ and $\tilde{\bf a}_y = -{\bf a}_x + {\bf a}_y$.

An interesting observation is that the magnetic configuration in Fig. \ref{Fig: Corner modes schematic} (a) is antiferromagnetic along $\tilde{\bf a}_x$ direction but ferromagnetic (FM) along $\tilde{\bf a}_y$ direction. To clarify the competition between AFM and superconductivity, Zhang {\it et al.} chose to first turn off s-wave pairing and the remaining system can be viewed as a 2d TI with additional AFM ordering. On the FM edge along $\tilde{\bf a}_y$ direction, the TI edge states will develop a Zeeman gap due to the FM exchange coupling effect. In spite of the explicit TRS breaking on the AFM edge along $\tilde{\bf a}_x$ direction, there exists an AFM TRS symmetry that ensures the gaplessness of the TI edge states, which consists of conventional TRS operation and a half-unit-cell translation along $\tilde{\bf a}_x$.

\begin{figure}[t]
	\centering
	\includegraphics[width=1.0\textwidth]{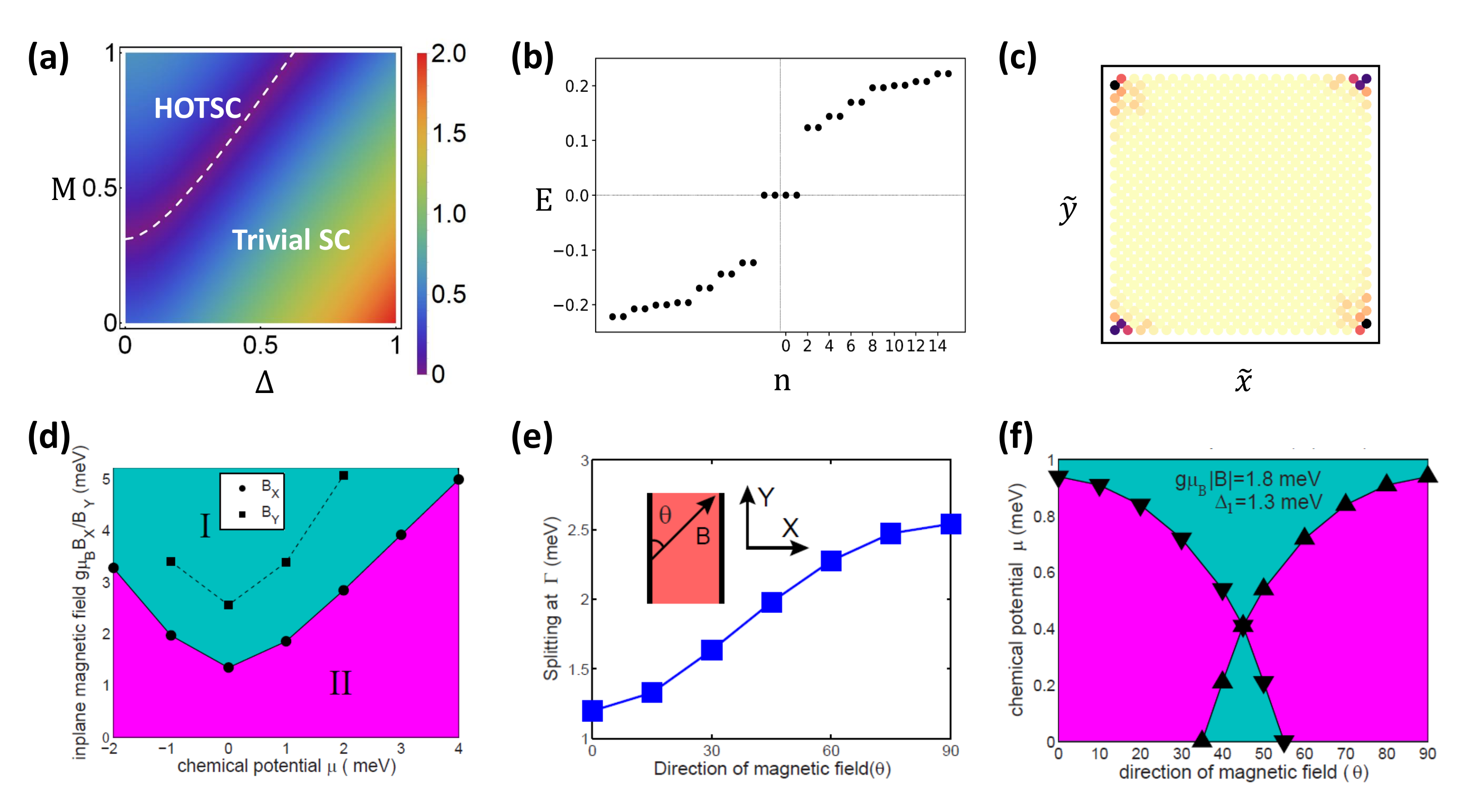}
	\caption{ Fig. (a) - (c) and Fig. (d) - (f) discuss the corner Majorana physics for the proposals from Zhang et al. \cite{zhang2019higher} and Wu et al.\cite{wu2019high}, respectively. (a) Topological phase diagram for a fixed $\mu=0.2$. The white dashed line shows the analytical results of Eq. \ref{Eq: Topo condition for corner modes}. (b) Energy spectrum of HFTS with open boundary conditions in both $\tilde{\bf a}_x$ and $\tilde{\bf a}_y$ directions, which clearly reveals four Majorana zero modes. (c) Spatial profile of the Majorana zero modes in (b).  (d) topological phase transition as a function of magnetic fields and chemical potentials for the (100) edge. The black circles (squares)
correspond to the TPT line for the magnetic field $B_X$ ($B_Y$)
perpendicular (parallel) to the edge. (e) Zeeman splitting of edge states as a function of magnetic field angle
with respect to 1d edge (Y axis), where $g\mu_B|B|$ is fixed to
be 5 meV. Inset shows the angle between magnetic field and
1d edge (thick black line). (f) Phase diagram of the existence
regime for MZMs at the corner as function of chemical potential $\mu$ and magnetic field angle $\theta$. In the pink/blue regimes,
the helical edge states at two perpendicular edges have different/the same topological characters, thus can/cannot host
MZMs at the corner. }
	\label{Fig: Corner modes}
\end{figure}

When superconductivity is turned on, the AFM edge immediately opens up a pairing gap. The higher-order topology emerges when the s-wave pairing fails to compete with the Zeeman gap on the FM edge. If this happens, the corner between AFM edge and FM edge will form a 0d domain wall between FM and superconducting gaps for TI edge states. As shown in Ref. \cite{fu2009josephson}, such domain wall will necessarily bind a single localized Majorana zero mode. Mathematically, Zhang {\it et al.} formulated a simple effective edge theory for the FM edge and analytically identified the condition for corner Majorana physics, which is given by
\begin{equation}
	M>\frac{1}{\beta_M}\sqrt{\mu^2+\Delta^2}.
	\label{Eq: Topo condition for corner modes}
\end{equation}
Here $\mu$ is the chemical potential and $\beta_M$ is a non-universal constant that depends on the details of band parameters. In Fig. \ref{Fig: Corner modes} (a), the white dashed line shows the analytical prediction of the topological phase boundary, which agrees well with the numerical results by calculating the FM edge spectrum. When placed on an open boundary geometry with the size of $20\tilde{a}_y\times 10\tilde{a}_x$, as shown in Fig. \ref{Fig: Corner modes} (b), four zero-energy Majorana modes are found inside both the bulk and the edge gaps, with each of them living on one of the four system corners [see Fig. \ref{Fig: Corner modes} (c)]. The emergent corner-localized Majorana zero modes unambiguously establish the FTS/FeTe bilayer heterostructure as a 2d higher-order TSC. Zhang {\it et al.} also made estimates on the important energy scales involved, which are based on known experimental results and first-principles calculations. In particular, they found the pairing $\Delta$ usually less than 10 meV \cite{zhang2016superconducting} and a fairly large exchange gap (around 60 meV) induced in the FTS layer. Given $\beta_M\approx 0.5$ and a small chemical potential $\mu$, the corner Majorana physics is experimentally accessible, especially since the growth technique is mature and has been successfully applied to achieve similar heterostructures in iron chalcogenides \cite{sun2014high,Nabeshima2017growth}. \\

The key ingredient for higher-order topology in the FTS/FeTe heterostructure is the competing gaps at two orthogonal edges (the magnetic gap at the FM edge and the SC gap at the AFM edge). Similar idea can also be applied to a monolayer FTS under an in-plane magnetic field (Fig. \ref{Fig: Corner modes schematic}(c)). Without magnetic field, the helical edge state is expected to open a SC gap while the Zeeman coupling can induce a magnetic gap for a large enough magnetic field. To support this competing edge gap picture, Wu {\it et al.} numerically studied energy spectrum along the 1d edge of FTS based on both the tight-binding model and the effective BHZ model \cite{wu2019high}. Fig. \ref{Fig: Corner modes}(d) shows the phase diagram of the helical edge states as a function of in-plane magnetic field and chemical potential for a fixed pairing gap. An edge phase transition is found and separates two topologically distinct phases, one with magnetic gap and the other with SC gap. However, to obtain higher-order topology, there is a crucial difference for this case compared to the FTS/FeTe heterostructure. In the FTS/FeTe heterostructure, the magnetic gap induced by exchange coupling only exists at the FM edge, but not the AFM edge, while for the in-plane magnetic field, the Zeeman coupling generally exists for the helical edge states along all the edges. While the s-wave SC gap is generally isotropic for any edge, higher-order topology {\it only} appear if the Zeeman coupling is {\it anisotropic} for the edge states along different edges. Therefore, Wu {\it et al.} numerically studied the magnetic gap as a function of the angle $\theta$ between the edge and in-plane magnetic field, and found a significant difference in the magnetic gap (about 80$\%$ in Fig. \ref{Fig: Corner modes}(e)) when the in-plane magnetic field is parallel or perpendicular to the edge.
Due to the anisotropic Zeeman coupling, a unique feature of BHZ model, the topological phase transition lines along one edge for different magnetic field directions (or equivalently for different directional edges with a fixed magnetic field) can be well separated, as shown by the solid and dashed lines for magnetic fields perpendicular or parallel to the edge in Fig. \ref{Fig: Corner modes}(d) and by up-pointing and down-pointing triangles for two orthogonal edges respectively as a function of the rotating angle $\theta$ of the in-plane magnetic field in Fig. \ref{Fig: Corner modes}(f). In Fig. \ref{Fig: Corner modes}(f), topological properties of two orthogonal edges are the same (distinct) in the blue (pink) regions. Therefore, the MZMs at the corner are expected when $\mu$ is tuned to the pink region.

The direct calculation of the system with the open boundaries can verify the existence of the corner MZMs, exactly the same as Fig. \ref{Fig: Corner modes}(b) and (c) for the FTS/FeTe heterostructure. Besides the corner MZMs, Wu {\it et al.} also predicted that by patterning local electric gates, the MZMs can also be found at the domain wall of chemical potentials at one edge and certain type of tri-junction in the 2d bulk, as depicted in Fig. \ref{Fig: Corner modes schematic}(c).
Given the high $T_c$ of 40 K \cite{Li2015PRB} and a large in-plane upper critical magnetic field of about 45 T \cite{Salamon2016}, the proposed setup is quite feasible in experiments. Compared to the heterostructure with the fixed AFM direction, utilizing the in-plane magnetic field has the advantage that it can be rotated and thus may provide an efficient approach to perform the braiding operation for the corner MZMs.

\section{Discussion and Perspective}\label{sec:perspective}
Recently, 112 family of iron pnictides have been predicted to host MZMs owing to the intrinsic TI-SC hetero-structure\cite{WuXX2020}. Interestingly, the systems are in a boundary-obstructed topological superconducting phase driven by the $s_{\pm}$ pairing. The experimental detection of MZMs will provide a smoking gun evidence for the extended $s$-wave pairing in the iron pnictides\cite{WuXX2020}. Besides the proposals based on topological band structures, recent STM experiments identified a zero-bias peak at the ends of atomic line defects in monolayer Fe(Te,Se)/STO, highly resembling the characteristics of MZMs\cite{ChenC2020}. They are theoretically interpreted as Majorana ends states of topological Shockley defects\cite{ZhangY2020} or topological magnetic defects\cite{WuXX2020-2}, waiting for further experimental verifications.

The above discussions have theoretically demonstrated the potential realization of Majorana modes at different locations, including the vortex core at the surface, the hinge of a 3d sample, the corner of a 2d sample, and the chemical potential domain walls, in iron-based SCs (monolayer, bulk or heterostructure). However, compared to the heterostructure approach combining the conventional SCs and semiconductors or TIs, both the experimental and theoretical studies in iron-based SCs are still at the infancy stage.

Current experiments in topological aspect of iron-based SCs mainly focus on several 3d bulk iron-based SC compounds, including FTS \cite{Zhang2018}, Li(Fe,Co)As \cite{Zhang2019NP}, (Li$_{1-x}$Fe$_x$)OHFeSe \cite{PhysRevX.8.041056} and CaKFe$_4$As$_4$ \cite{2019arXiv190700904L}. As discussed above, the topological electronic band structure and topological surface states in these compounds have been well established through high-resolution spin-resolved ARPES measurements. The zero bias peak at the vortex core has been observed at the surface of FTS and (Li$_{1-x}$Fe$_x$)OHFeSe through the STM measurements, providing evidence of MZMs \cite{PhysRevX.8.041056}. However, it turns out that the selective appearance of zero bias peaks (around 20\% vortex cores during more than 150 measurements) challenges the Majorana interpretation, and its physical origin is still under debate. Unlike spin-resolved STM measurement in SC/TI heterostructure \cite{PhysRevLett.116.257003}, spin information of the zero-bias peak has not been extracted yet. A more recent STM study claims to observe a peak value nearly reaching $\frac{2e^2}{h}$ after removing extrinsic instrumental broadening via deconvolution \cite{2019arXiv190406124Z}. However, just like the early stage experiments in the heterostructure approach \cite{lutchyn2018majorana}, a large sub-gap density of states still coexists with the zero-bias feature (soft-gap), and these sub-gap states are detrimental to the topological protection of the coherence of MZMs. Therefore, the next material challenge in iron-based SCs is to improve sample quality and remove these sub-gap states.

For 2d monolayer FTS, it is still challenging to confirm the topological electronic band structure. As discussed above, a recent ARPES experiment shows the evidence of the gap closing when tuning the ratio between Te and Se compositions, which is supported by the first principle calculations.\cite{Peng2019} The STM measurements also show additional density of states at the edge, as compared to that in the bulk, which is attributed to the 1d helical edge state. However, unlike the other QSH systems with the characteristic $\frac{2e^2}{h}$ two-terminal conductance, the existence of 2d bulk Fermi pockets and the high SC transition temperature in FTS prevents the direct observation of the quantized conductance through transport measurements. Therefore, it is still an open question about how to unambiguously demonstrate the helical edge physics in this system. Compared to the 3d case, the 2d monolayer FTS has the advantage of higher $T_c$, higher upper in-plane critical field and better tunability. Patterning local electric gates or rotating magnetic fields will allow for the manipulation of MZMs and thus may provide a variable route towards topological qubit operations via the braiding and fusion of MZMs.
In addition, the phase difference between two SC regions in a Josephson junction provides an additional control knob for tuning Majorana modes. Indeed, recent experimental efforts have demonstrated the feasibility of creating MZMs at the end of a Josephson junction in the semiconductor/SC hybrid system \cite{RenHC2019,FornieriA2019}. Such approach has not been explored in iron-based SCs.

In addition to superconductivity, iron-based compounds also exhibit rich phase diagram with other phases, including anti-ferromagnetism and nematicity \cite{RevModPhys.87.855,Fernandes2014}. A large number of research works focus on the competition between superconductivity, anti-ferromagnetism and nematicity, while their influence on topological states is much less understood. Interestingly, it has been theoretically suggested that nematicity can drive a topological phase transition and lead to the helical edge states in FeSe \cite{2016arXiv160302055W}. Given the rich phase diagram, topological defects, such as domain wall, dislocation, vortex etc., may also exist, and thus iron-based SCs also provide a fertile platform to explore the interplay between topological electronic band structure and topological defects.

\section*{Acknowledgments}
C.X.L acknowledges the support of the Office of Naval Research (Grant No. N00014-18-1-2793), the U.S. Department of Energy (Grant No.~DESC0019064) and Kaufman New Initiative research grant (grant No. KA2018-98553) of the Pittsburgh Foundation. R. X. Z is supported by a JQI Postdoctoral Fellowship. G. X is supported by the Ministry of Science and Technology of China (No. 2018YFA0307000) and the National Natural Science Foundation of China (No. 11874022). J.P.Hu was supported by the Ministry of Science and Technology of China 973 program
(No. 2017YFA0303100), National
Science Foundation of China (Grant No. NSFC11888101), and the
Strategic Priority Research Program of CAS (Grant
No.XDB28000000).

\bibliographystyle{ws-rv-van}
\bibliography{FeTeSe_TSC-review_final}

\end{document}